# Calm Before the Storm: The Challenges of Cloud Computing in Digital Forensics

GEORGE GRISPOS   TIM STORER   WILLIAM BRADLEY GLISSON

## ABSTRACT

Cloud computing is a rapidly evolving information technology (IT) phenomenon. Rather than procure, deploy and manage a physical IT infrastructure to host their software applications, organizations are increasingly deploying their infrastructure into remote, virtualized environments, often hosted and managed by third parties. This development has significant implications for digital forensic investigators, equipment vendors, law enforcement, as well as corporate compliance and audit departments (among others). Much of digital forensic practice assumes careful control and management of IT assets (particularly data storage) during the conduct of an investigation. This paper summarises the key aspects of cloud computing and analyses how established digital forensic procedures will be invalidated in this new environment. Several new research challenges addressing this changing context are also identified and discussed.

## INTRODUCTION

Cloud computing technologies have significant potential to revolutionise the way organizations provision their information technology (IT) infrastructure. Migration to cloud computing involves replacing much of the traditional IT hardware found in an organization's data centre (including servers, racks, network switches and air conditioning units) with virtualized, remote, on-demand software services, configured for the particular needs of the organization. These services can be hosted and managed by the user organization (on a reduced hardware base), or by a third-party provider. Consequently, the software and data comprising the organization's application may be physically stored across many different locations, potentially with a wide geographic distribution.

There have been several predictions of substantial market growth in cloud services over the next few years. Gens has speculated that spending on cloud services will grow by 30% in 2011 (Gens, 2010). A Gartner press release forecast cloud service worldwide revenue to reach $68.3 billion in 2010, an increase of 16.6% from the 2009 revenue of $58.6 billion, and goes on to claim that cloud service revenues will reach $148.8 billion in 2014 (Pring, Brown, Leong, Biscotti, Couture, Lheureux, Frank, Roster, Cournoyer, & Liu, 2010). A study at the end of 2010 predicted that within the next three years, approximately 40% of Small and Medium Businesses (SMBs) expect to be using three or more cloud services and will have migrated their data into the cloud (Kazarian & Hanlon, 2011). There is some speculation that new and SMBs will benefit the most in the coming years, with cloud computing allowing these organizations to utilize

appropriately scaled IT infrastructure that was previously only accessible to larger corporations (Schubert, Jeffery, & Neidecker-Lutz, 2010).

The use of cloud computing has potential benefits to organizations, including increased flexibility and efficiency. Virtualized services provide greater flexibility over an in-house physical IT infrastructure, because services can be rapidly re-configured or scaled to meet new and evolving requirements without the need to acquire new and potentially redundant hardware. Complementary to this, the use of cloud computing can reduce the costs of providing IT services, by eliminating redundant computing power and storage, reducing support requirements and reducing fixed capital commitments. Khajeh-Hosseini et al. found that a 37% cost saving could be obtained by an organization who chose to migrate their IT infrastructure from an outsourced data-centre to the Amazon Cloud (Khajeh-Hosseini, Greenwood, & Sommerville, 2010).

However, the use of cloud computing presents significant challenges to the users of clouds (both individuals and organizations), as well as regulatory and law enforcement authorities. It has been estimated that cybercrime will cost the British economy £27 billion per year in the coming years, with businesses accounting for nearly £21 billion of losses, largely due to the theft of intellectual property and espionage (Detica, 2011). It is likely that users of cloud computing services and technologies will be exposed to similar risks. The security of confidential corporate and private data remains one of the greatest concerns organizations have when they consider cloud computing (Butler, Heckman, & Thorp, 2010). Recent reports have noted Botnet attacks on Amazon's cloud infrastructure (Amazon Web Services, 2009). The compromise of the Gmail email service by (alleged) Chinese hackers (Blumenthal, 2010) illustrates that cloud computing platforms are already a target for malicious activities.

When security breaches, attacks or policy violations occur, it may be necessary to conduct a digital forensic investigation. However, existing digital forensic principles, frameworks, practices and tools are largely intended for off-line investigation. In particular, these approaches assume that the storage media under investigation is completely within the control of the investigator. Conducting investigations in a cloud computing environment presents new challenges, since evidence is likely to be ephemeral and stored on media beyond the immediate control of an investigator. This paper raises the awareness on the challenges posed by cloud computing technologies for digital forensics through an analysis of the applicability of the Association of Chief Police Officers (ACPO) digital forensics principles and the Digital Forensics Research Conference (DFRW) Investigative Process Model (DIP Model) to a cloud computing context. These two frameworks are commonly cited as the basis of good digital forensic practice (Hunton, 2011; Ieong, 2006; Owen & Thomas, 2011; Reyes, O'Shea, Steele, Hansen, Jean, & Ralph, 2007). The paper identifies numerous aspects of these approaches that are problematic in a cloud context and proposes several immediate areas of research that the authors believe is fundamental to further understand digital forensic investigations in the cloud.

This paper is structured as follows. Section 2 summarises the aspects of cloud computing technologies that are pertinent to a digital forensic investigation. Section 3 reviews existing digital forensic process models, and notes the key assumptions made regarding the environment under investigation and Section 4 assesses how the adoption of cloud computing may invalidate the assumptions made in these models. A review of related work is presented in Section 5, illustrating how some of these challenges concerning cloud computing which we identify are beginning to be addressed. Section 6 discusses some further avenues of research to address the issues raised by the analysis. Finally, Section 7 draws some conclusions and summarises the key challenges in conducting digital forensic investigations in cloud environments.

## CLOUD COMPUTING

A cloud has several uses, offering a variety of services and can be deployed in more than one way. Consequently, several definitions of cloud computing have been proposed (Mell & Grance, 2011; Schubert, et al., 2010; Wyld, 2009).

Schubert et al. define cloud computing as:

> *a 'cloud' is an elastic execution environment of resources involving multiple stakeholders and providing a metered service at multiple granularities for a specified level of quality (of service).* (Schubert, et al., 2010)

The Open Cloud Manifesto Consortium defines the key aspects of cloud computing as:

> *the ability to scale and provision computing power dynamically in a cost-efficient way and the ability of the consumer (end user, organization, or IT staff) to make the most of that power without having to manage the underlying complexity of the technology.* (The Open Cloud Manifesto Consortium, 2009).

The key aspects from a digital forensic perspective is that a cloud is defined as a scalable, virtualized, distributed computing platform, whose shared resources are accessed remotely by users through a network.

There are three main levels of service for users of cloud computing (Mell & Grance, 2011):

- in the *Software as a Service* (SaaS) model, a client can make use of software applications made available from the cloud provider. Typically, users interact with SaaS applications using a web-browser. An example of SaaS is the Google Apps[i] suite offered by Google. Clients can use this service to deploy an email and collaboration platform within their organizations, and make use of Google Docs, Calendar, Gmail and other productivity applications. All data generated by the use of the applications is stored in the cloud;

- the *Platform as a Service* (PaaS) model provides an application programming interface (API) for clients to create and host custom-built applications. An example of PaaS is the Google App Engine[ii], which provides a platform for developers to create and host web-based applications. PaaS also includes cloud providers offering database management systems such as Amazon SimpleDB[iii]; and

- the *Infrastructure as a Service* (IaaS) model is the leasing of virtualized computing resources such as processing power, volatile memory and persistent storage space to host virtual machines. IaaS products include Amazon EC2[iv], which allows clients to create and launch virtual machines running a variety of operating systems. These can then be loaded with customer-specified applications, just as for any other server. The virtual machine image can be stored and re-deployed, according to the client's requirements.

The manner in which services are deployed in a cloud can influence the evidence available to an investigator and the way it is collected. For example, IaaS platforms present an interface to a user that is indistinguishable from that of a remote physical server. However, the data that represents an IaaS-based server is inherently more volatile. Alternatively, the SaaS and PaaS models restrict the flexibility with which users can interact with a cloud platform, by offering a restricted set of applications, or specifying the constraints within which new software can be created. The storage of data on these services is not by the user, but instead by the cloud owner.

In addition to the different levels of deployment, a cloud can be categorised by its organizational deployment, with consequent impact on the geographical location and storage architecture of data held (Krutz & Vines, 2010):

- in a *private* cloud, the infrastructure is operated solely by the organization who owns the cloud. This cloud will likely be found within the same premises as the owning organization and be within its administrative control, and include only that same organization's data;

- a *community* cloud is shared between several organizations, either because of a common organizational goal, or in order to pool IT resources. Community clouds may be located within one or more of the community organization's premises, and will be administered by the community;

- *public* clouds will usually be owned by a provider organization, which will maintain the cloud facilities in one or more corporate data centres. The administrative control of the cloud resources will therefore reside with the provider, rather than the user. Consumers will lease virtual storage and compute resources from the provider as required. A public cloud will therefore likely contain data from more than one user; and

- a *hybrid* cloud is a composition of two or more of the above deployment options. Hybrid clouds can be used to provide load balancing to multiple clouds. For example, an organization may have exhausted the available resources within its private cloud, and so incorporate resources available on lease from a public cloud.

A consequence of these different organizational configurations may have an impact on the way that data can be collected as evidence. In particular, the data held in a cloud may be physically stored in one or more geographically distributed locations, making the determination of which legal framework and procedures to apply to the evidence gathering process difficult.
In summary, multiple deployment options and the variety of services offered to cloud users introduce new challenges when conducting digital forensic investigations in these environments. The next section of this paper summarises various models used in digital forensic investigations.

# CURRENT DIGITAL FORENSIC PROCESS MODELS

The Association of Chief Police Officers (Association of Chief Police Officers & 7Safe, 2007) propose four principles of digital forensic practice:

1. No action taken by law enforcement agencies or their agents should change data held on a computer or storage media, which may subsequently be relied upon in court.
2. In circumstances where a person finds it necessary to access original data held on a computer or on storage media, that person must be competent to do so and be able to give evidence explaining the relevance and the implications of their actions.
3. An audit trail or other record of all processes applied to computer-based electronic evidence should be created and preserved. An independent third party should be able to examine those processes and achieve the same result.
4. The person in-charge of the investigation (the case officer) has overall responsibility for ensuring that the law and these principles are adhered to.

These guidelines are primarily intended for law enforcement investigators, but have also been adopted by the digital forensics community in the United Kingdom (UK). There is an expectation that evidence used in courts produced from a digital forensic investigation will be gathered by following these guidelines (although it is unclear how this determination is made in practice).

Since 2001, various frameworks and process models have also been proposed for conducting a digital forensics investigation (Baryamureeba & Tushabe, 2004; Carrier & Spafford, 2003, 2004; Reith, Carr, & Gunsch, 2002). The First Digital Forensics Research Conference (DFRW) defined the term 'digital forensics' and proposed the DFRW Investigative Process (DIP) Model, which they deemed could be applied to all investigations in both research and practitioners (Palmer, 2001). This was the first attempt to define how a digital forensics investigation should be conducted. The model identifies a linear process, which includes stages of: *identification, preservation, collection, examination, analysis and presentation*. As Palmer (2001) noted, the conference agreed that this model still required further work and as such was not deemed complete.

The Abstract Digital Forensics Model builds on the work from the DFRW and is effectively an expansion of the DIP Model (Reith, et al., 2002). It adds *preparation, approach strategy* and *returning evidence* phases to the DIP Model, as well as describing the phases, something that the DIP Model lacked.

Carrier and Spafford (2003) proposed the Integrated Digital Investigation Process, based upon "theories and techniques from the physical investigation world". This model is based upon on 'physical crime scene' principles and methods and consists of 17 phases that are broken into five groups: *readiness, deployment, physical crime scene investigation, digital crime scene investigation* and *review*.

Finally, The Enhanced Digital Investigation Process Model is an enhancement of the model proposed by Carrier and Spafford (2003), consisting of five major phases (Baryamureeba & Tushabe, 2004):

- the *readiness* phases is concerned with ensuring the investigator has the correct training and infrastructure to handle an investigation;

- the *deployment* phases specifies means for detecting an incident has occurred and beginning the process of conducting an investigation;

- during the *traceback* phases the crime scene is examined and the devices worthy of investigation are discovered;

- the *dynamite* phases is associated with collecting and analysing items seized from the crime scene and collecting evidence from these devices; and

- the *review* phases is concerned with reviewing the entire investigation and discovering areas of improvement.

All of these guidelines were developed prior to the advent of cloud technologies and largely assume that the investigator has physical access and control over the target system or device, and in particular, its storage media. This assumption is likely to be invalidated when investigating activity in a cloud environment.

## DIGITAL FORENSICS IN CLOUD ENVIRONMENTS

This section analyses the challenges raised by cloud computing with respect to existing models of digital forensic investigations described above. The following discussion is based principally on the DFRW Investigative Process (DIP) Model and the ACPO principles and guidelines. The DIP model provides a comprehensive review of the stages employed in the digital forensic process, and so is convenient for analysing the impact of cloud forensics on this process. Other models are also referenced where appropriate, in particular the ACPO principles and guidelines.

The issues raised concerning cloud computing in each of the phases of a digital forensic investigation are summarised in Table 1. The analysis demonstrates that many of the assumptions incorporated into existing models of forensic investigation are not valid in the context of cloud computing.

### Identification

The first step of the DIP model is the determination that a potential criminal or improper act has taken place involving computer-based systems. These events may relate to traditional crimes or activity augmented by the use of IT, or IT-specific crimes. Identification may result from, for example, complaints made by individuals, anomalies detected by Intrusion Detection Systems (IDS), monitoring and profiling or because of an audit of a computer system. Although the identification phase is not just concerned with digital forensics, it does have an impact on how the investigation is conducted as well as defining the purpose for conducting the investigation.

The detection of suspicious events in a cloud will depend on the deployment model adopted and the form of cloud services (SaaS, PaaS or IaaS) used. The deployment of conventional intrusion detection systems in a cloud has been proposed by several authors (Roschke, Cheng, &

Meinel, 2009; Vieira, Schulter, Westphall, & Westphall, 2010). Such systems could be deployed by users of IaaS clouds, or by providers in SaaS or PaaS clouds. In a private cloud infrastructure, providers may be better placed to tune the IDS for the particular suite of services deployed which meets the organization's needs. For public clouds, a multi-layered strategy may be necessary. Users can monitor for suspicious events occurring with the services they are using. Providers can monitor the underlying infrastructure used to host the cloud, and therefore detect much larger attacks that could affect a much larger audience.

## Preservation and Collection

A digital forensic investigation is concerned with collecting data from computer-based systems that can later be constituted as evidence that a crime or other illicit act has been committed. Legal convention and forensic standards, such as the Daubert principles (Marsico, 2004), require that forensic evidence be testable, and that the methods used to produce evidence be repeatable. Consequently, the preservation phase of the DIP model defines activities prior to data collection to ensure the integrity of data throughout the investigation life cycle, i.e. assurance that the evidence is an accurate representation of the data found on the computer system. Several aspects of the preservation phase are affected by the use of a cloud environment.

### *Storage Capacity*

In conventional investigations, a pre-requisite of evidence preservation is to have available sufficient secure storage capacity for the data gathered to be archived. Several authors have noted that the growing amounts of data gathered during forensic investigations, driven by increased device capacity and reduced cost, is increasingly challenging for investigators (Richard & Roussev, 2006; Roussev, Wang, Richard, & Marziale, 2009; Sommer, 2004). This increase imposes extra costs on investigators with the responsibility to store and curate the data, quite apart from the increasing amount of investigator time required to examine it.

The use of cloud environments will likely exacerbate the problem of data storage. An attractive aspect of cloud environments for users is the *elastic* ability to dynamically scale a service's storage capabilities according to on-going requirements. From a user's perspective, a typical public IaaS cloud appears to offer limitless data storage capability as and when the user requires it. An investigator may be faced with gathering an extremely large amount of data placed in a cloud by a user.

One solution investigating authorities could resort to is the use of public clouds to store evidence. This too will bring its own challenges, from both a legal and technical perspective. Investigators will need to address the rules and regulations regarding data protection and privacy issues, and their impact on evidence stored in the cloud.

The adoption of triaging techniques has also been proposed as a means of reducing the amount of data to be analysed by an investigator (Pearson & Watson, 2010) and is already being adopted to reduce backlogs in conventional investigations (Rogers, Goldman, Mislan, & Wedge, 2006). This approach may be particularly appropriate in situations that require urgent responses, such as kidnappings, when the long-term integrity and reliability of evidence is less important.

| Phase | Action | Challenges |
|---|---|---|
| *Identification* | Identifying an illicit event | Lack of frameworks |
| *Preservation* | Software tools | Lack of specialist tools |
| | Sufficient storage capacity | Distributed, virtualized and volatile storage; use of cloud services to store evidence |
| | Chain of custody | Cross-jurisdictional standards, procedures; proprietary technology |
| | Media imaging | Imaging all physical media in a cloud is impractical; partial imaging may face legal challenges |
| | Time synchronization | Evidence from multiple time zones |
| | Legal authority | Data stored in multiple jurisdictions; limited access to physical media |
| | Approved methods, software and hardware | Lack of evaluation, certification generally, but particularly in cloud context |
| | Live vs. Dead acquisitions | Acquisition of physical media from providers is cumbersome, onerous and time consuming data is inherently volatile |
| | Data integrity | Lack of write-blocking or enforced persistence mechanisms for cloud services and data |
| *Examination* | Software tools | Lack of tested and certified tools |
| | Recovery of deleted data | Privacy regulations and mechanisms implemented by providers |
| | Traceability and event reconstruction | Events may occur on many different platforms |
| *Presentation* | Documentation of evidence | Integration of multiple evidence sources in record |
| | Testimony | Complexity of explaining cloud technology to jury |

*Table 1: Summary of Challenges to Digital Forensics in Cloud Environments*

The proposed approach permits investigators to conduct examinations of storage devices in a short period in order to identify the most valuable evidence without performing a full forensic investigation.

The Computer Forensics Field Triage Process Model (CFFTPM) is a framework proposed to perform triage on digital devices (Rogers, et al., 2006). CFFTPM works by accessing information located in a user's 'home directory' on the file system, which contains user-application centric information. Other sources of information used in the CFFTPM include the registry, operating system and application logs, all of which can contain timestamp information.

The CFFTPM will likely not transfer directly to the context of a cloud environment, since user-centric application data may be stored in the cloud, cached on the user's client computer, or both. Adopting a triage approach may require an investigator to conduct a live examination of this data in the cloud environment while the client is still connected. The implications of a live investigation are discussed further below in the context of data acquisition.

### *Chain of Custody*

During a conventional forensic investigation, accepted practice is to establish and maintain a 'chain of custody' for evidence, which is defined as:

> *a roadmap that shows how evidence was collected, analysed and preserved in order to be presented as evidence in court.* (Vacca, 2005)

A properly maintained chain of custody therefore provides the documentary history for the entire lifetime of evidence discovered during an investigation. ACPO guidelines stipulate that the documentation include how the evidence was gathered and managed, by whom and when.

In a conventional investigation, the chain of custody begins when an investigator assumes physical control of digital electronic artefacts (and any incorporated storage devices) that is suspected to be pertinent to the investigation. Subsequently, there are two methods of preserving data on a personal computer (Kruse, Warren, & Jay, 2001): powering down the computer by issuing a command to the operating system causing a staged shutdown, and removing the power source, causing an immediate halt. Storage devices can then be removed from the computer and examined separately. The chain of custody documentation will typically refer to these devices, which can be isolated and disconnected from a power supply with little risk of loss of evidence.

The remote nature of cloud services means that this assumption is not valid in the context of a cloud environment. Services can be accessed by any system with a network connection to the hosting cloud. Unless an investigator is able to gain control of and disable a service, evidence could be destroyed relatively quickly, either by a service user, or by the cloud provider. To the authors' knowledge, there has been little work by either researchers or practitioners to examine the practicality of obtaining control of a cloud service during an on-going forensic investigation. Challenges in this context include the speed with which an investigator can gain control of a service, and the appropriate legal and regulatory framework that should be developed to enable this capability.

## *Digital Image Acquisition*

Assuming that the investigator has gained control of the cloud service it is necessary to obtain an accurate copy of the data held by the service for later analysis. Both the DIP Model and ACPO guidelines assume the use of 'forensic imaging' to obtain copies of a storage device's contents without alteration of the source (Association of Chief Police Officers & 7Safe, 2007; Palmer, 2001). Typically, a storage device is connected to an investigator's own computer via a write-blocker as shown in *Figure 1a*. A byte-for-byte copy of the entire device (an image) is then made using a software tool such as AccessData's FTK Imager[v] or the open-source tool *dd*[vi]. If multiple copies of the image are taken, digital hashes of each image can be taken to check whether the source image has been changed.

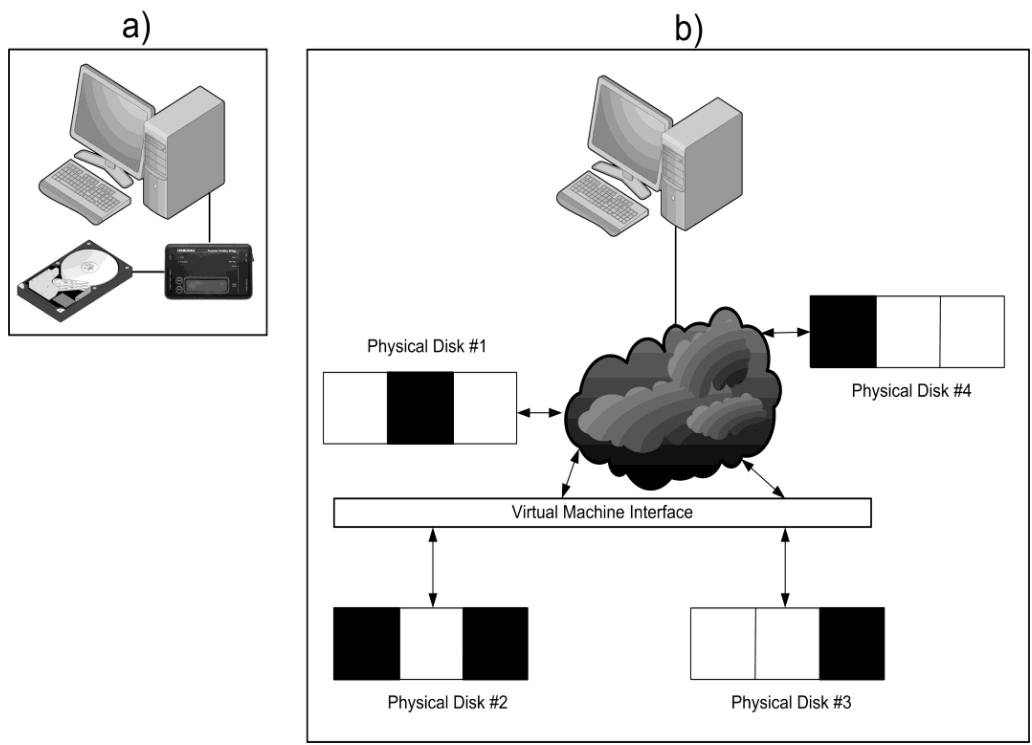

*Figure 1. Digital image acquisition in traditional (a) and cloud computing environments (b)*

The collection of evidence from a cloud environment is likely to pose a challenge to investigators. Triage tools, volatile and persistent memory acquisition software, as used in conventional investigations, on a client computer may provide minimal data.

The virtualization of data storage in a cloud makes it complex to identify and isolate the portions of the one or more physical storage devices owned by a cloud provider that represent the user's data that should be gathered for analysis. Virtualized data stored on a cloud may be spread between many different physical devices and an interface between the virtual storage and the investigator could exist (*Figure 1b*). For example, Google have employed the Google File System (GFS) to store their customer's data in the cloud (Ghemawat, Gobioff, & Leung, 2003). To customers, data appears to be stored in a single location; however, physically this is not the case. GFS is a "multi-tenant distributed" file system which means that even if two users are within the same organization, their data could well reside in two or more different physical locations (Google, 2010).

The ACPO guidelines envisage the entire storage device containing relevant information to be collected (Association of Chief Police Officers & 7Safe, 2007). Acquiring all such storage devices from a cloud environment could be both cumbersome and time consuming for the investigator, and disruptive for the provider. The amount of data collected could also be very large, particularly in relation to the amount of relevant data contained. Fundamentally, cloud services only offer remote access to a *logical* representation of data, rather than the underlying *physical* infrastructure. This limitation is likely to be complicated further by the provision of cloud services whose infrastructure is itself virtualized and leased from other cloud providers (Bessani, Correia, Quaresma, André, & Sousa, 2011). Logistically, it appears inevitable that methods will be needed which allow for only a partial recovery of data from any one physical device. Such a method would need to be developed in accordance with accepted forensic principles. In particular, the risk of conducting an investigation with incomplete data must be addressed (Carrier & Spafford, 2004).

This use of virtualization also affects the privacy of other users of the cloud, whose data may be inadvertently gathered during the investigation. In some jurisdictions, inadvertent access of non-relevant data from a cloud environment may contravene local privacy and/or data protection legislation.

The preceding discussion has assumed that the investigator conducts a 'dead acquisition' on a physically isolated storage device. However, frequently used data may be stored in volatile memory on the cloud, or be cached by a user's computer during interactions with cloud services. 'Live' acquisitions and investigations are an alternative approach, in which data is examined on the target computer while it is still powered up. This approach enables investigators to gather data that might otherwise be lost if a computer is powered down, particularly:

- data stored in non-persistent memory, such as processes and information on active network connections; and

- temporary data stored in persistent memory, such as application file locks, and web-browsing caches.

The use of live acquisition techniques may increase the amount of information an investigator is able to extract from a cloud client computer, particularly if it has an open connection to a cloud environment. However, digital image acquisition could be further hampered by the use of encryption in cloud-based environments. With many organizations reluctant to adopt cloud services and storage until concerns about data confidentiality and integrity is met (Kamara & Lauter, 2010), cloud service providers are turning towards encryption as a means of offering this security to their customers. Several cloud storage providers such as SpiderOak[vii] have implemented a 'zero knowledge system' such that all data is encrypted client-side before being transmitted and stored in the cloud, furthermore, the keys used to encrypt data are never stored in the cloud (Agudo, Nuñez, Giammatteo, Rizomiliotis, & Lambrinoudakis, 2011). This means serving a cloud storage provider with a court order to decrypt such information could prove fruitless as only the owner of the data can provide the key to decrypt this information. If further cloud providers deploy such an encryption system as a means assuring customers their data is safe, evidence in digital images created could reveal encrypted blocks of data of no forensic value to investigators unless encryption keys can be recovered.

### *Deleted Data*

The cloud could both assist and hamper investigators attempts to recover data that is deleted or would otherwise have been deleted by the suspect. Unlike a hard drive or USB flash drive, which a suspect has physical access and can therefore be physically destroyed; this is not the case with the cloud. Unless the suspect has the knowledge and administrative authority to delete or 'destroy' data, this evidence will remain available to the investigator.

In conventional investigations, data that a user has attempted to delete (but which still resides on a storage device) is often a rich source of evidence. However, the volatility and elasticity of cloud environments make the recovery of deleted data challenging. Some cloud providers maintain that user privacy is a priority within their cloud environments (Google, 2010). For example, Google's current policy regarding deleted data is such that once a user deletes their data from Google Services, that data is then deleted from both active and replication servers. Pointers to this data are also deleted, making tracing remnants of user deleted data extremely difficult.

The European Commission is encouraging European Union member states to implement the Data Retention Directive (European Union, May 2006). Article 5 of this directive requires member states to ensure that communication providers shall retain certain information about its users, including the "userID", "IP address allocated at the time of the communication" as well "the date and time of the log-in and log-off of the service" (European Union, May 2006). Cloud providers are not explicitly mentioned in the directive. However, should they be encompassed by the umbrella definition of 'communication providers', they too will also need to retain specific information related to clients.

### *Cross-Organizational Cooperation*

If it is not possible for an investigator to obtain personal control of a cloud service, it may be possible to obtain an image of the service's data from the cloud provider. However, this approach is problematic for several reasons. Using the cloud provider to obtain the image means that the investigator does not initially control the chain of custody documentation. A complete chain of custody should identify all individuals who have come in contact with the evidence. Consequently, if the cloud provider is used to obtain the initial image, then the chain of custody begins with the employees assigned to this task. The second principle in the ACPO guidelines states that the individual responsible for acquiring evidence must be competent to do so (although the guidelines are not explicit about what constitutes competency). It is unclear how an investigator would be satisfied that the cloud provider's employees were competent to gather evidence on their behalf.

The cross-organizational nature of this approach to evidence gathering from cloud environments has other implications. The investigator may have difficulty establishing the training and experience of the cloud providers employees assigned to assisting the investigation, particularly if the standards adopted within the provider organization are not directly comparable to those of the investigator. The use of proprietary technologies by cloud providers may make the involvement of the provider in the investigation essential. For example, parts of the GFS are considered proprietary business information. Investigators may require the cooperation of cloud providers to locate evidence. Consequently, the reliability and quality of the chain of custody may be insufficient for the investigator's purposes.

This problem is exacerbated if the cloud provider is hosted in a different jurisdiction to that of the investigator, since different legal norms and practices may govern the provider's behaviour. A local search warrant to search and seize evidence may not give the investigator the right to do the same with evidence located in different jurisdictions, even though the evidence is accessible via the Internet (K. Wang, 2010). For example, Amazon stores data from customers in the European Union in its cloud services in the Republic of Ireland (Wauters, 2008). This means that an investigator seeking evidence concerning a cloud user will need to work with the authorities in that jurisdiction. Such a request may take some time, by which time the evidence could be lost or deliberately destroyed.

A partial solution may be for cloud providers to have individuals within their organization, who are trained and qualified to perform forensic investigations should the need arise. These individuals can then begin a chain of custody, which will be passed onto the investigating party. However, it is unclear how the cloud provider will defray the costs of providing this service. It may be necessary for such services to be mandated by legislation.

The preservation phase also includes the reconciliation of timing information concerning digital evidence, in particular from time stamps in file system meta-data, and from application log files on one or more computer systems. This information can be used to re-construct the sequence of events concerning a suspicious event.

For an investigator to re-construct an accurate time-line of events on the device or system, the correct time and time zone need to be established. In a cloud environment, establishing this information could be challenging. Public clouds will potentially store evidence in a distributed manner across various physical locations. Hence, the physical locations could be in more than one time zone. In addition, virtualized services may also operate according to the time zones of their users, rather than their physically hosted locations.

## Examination and Analysis

Several examination techniques are discussed in the DIP Model once data has been preserved and collected, and there is a variety of software tools available to assist an investigator. Dedicated forensic tool suites such as Forensic Tool Kit[viii] or Encase[ix] are popular commercial choices. Sleuth Toolkit[x] is an open source alternative. These tool suites can be used to perform 'pattern matching' and 'filtering', which can involve either searching for specific filenames, file types, or content. These tool suites can also be used to discover and recover data that a user has attempted to delete.

During the analysis phase of an investigation, the significance of information artefacts as evidence is evaluated. A narrative is developed, supported by the evidence and a timeline to explain how a crime was committed. Where appropriate, it may be possible to associate particular artefacts with users or user accounts.

Evidence produced during analysis may also be subject to validation, either through comparison with complementary sources, or with previous versions, to gain assurance that the evidence has not be altered as a consequence of some analysis technique.

### *Types of Evidence in Clouds*

Many types of evidence found in clouds, will likely be similar to that found in conventional investigations, including Office application documents, emails and images. Several new forms of evidence will also be available, in particular records of activities of users with clouds. Major cloud providers such as Amazon and Google have implemented a number of logging mechanisms tracking use within their services:

- **Message Log Search** - is a service from Google, which allows administrators to make queries on email messages. Forensic investigators can also use this search provided they can gain access to the administrator account. Using this tool an investigator can find logs containing information such as: emails sent on a specific date, account ID identification for a specific email, identification of specific email recipients, and the IP address of the sending or receiving Mail Transfer Agent (Google, 2011).

- **Amazon Simple Storage Service (S3) Logging** - amongst other logging, Amazon provides logging for 'buckets' created using Amazon S3. Logging can be configured to record requests made against the bucket such as the request type, the resource which the request worked and the time and data the of the request (Amazon Web Services, 2010).

To access these logs, an investigator currently needs to access the administrative section of the cloud service under investigation. As such logs will be located in the cloud, the administrative username and password will be required to access them. Cloud providers could assist investigators, although the issues concerning chain of custody discussed above will be present. Marty has proposed a framework for recovering logging information during forensic investigations involving the cloud (Marty, 2011).

### *Validation using Hashing Tools*

Software hashing tools are commonly used in conventional investigations to validate the on-going integrity of data used as evidence. A hash function is an algorithm for converting arbitrary length data strings into fixed length *hash values*, typically a few hundred bytes in length. Hash functions are designed so that any change in the input data should (with high probability) produce a different output hash value. Hash values can therefore be periodically computed for disk images, files or other data representing forensic evidence to gain assurance that the evidence has not been changed by an analysis (Salgado, 2005).

Data stored in a cloud can also be subjected to hashing for integrity checking purposes. For example, Amazon S3 and Web Services (AWS) have both implemented MD5 hashing checksums for objects stored in their services (Amazon Web Services, 2010). In principle, an investigator can record these checksums to show that any evidence acquired has remained unchanged during the course of the investigation. In addition, this feature may be of future use for investigators wishing to store forensic evidence that they have gathered in a cloud environment.

The use of hashing tools implemented, deployed and controlled by cloud providers does raise some challenges. As in the discussion on forensic imaging above, the use of external facilities draws the provider into the chain of custody. In addition, the investigator has less opportunity to test and evaluate the hashing features in a cloud, compared with tools developed for use on conventional desktop PCs. Typically; an investigator can use a selection of tools that implement the same hash function to compute a hash for some sample data. Any differences between the results produced can be investigated. However, in a cloud environment, the investigator has only a single implementation (the checksum implementation deployed by the cloud provider) to use. Consequently, the investigator's ability to validate the correctness of their tools is limited.

## Presentation

Evidence gathered during a digital forensic investigation can be summarized to explain their conclusions in the form of a report or briefing. These may then be submitted to a court and an investigator could be asked to provide expert testimony and be subject to cross-examination (Carrier & Spafford, 2003). Alternatively, the results of an investigation could be used by an organization to improve their corporate policy and could evolve as a form of documentation for future investigations (Y. Wang, Cannady, & Rosenbluth, 2005).

In the United States and the United Kingdom (UK), expert scientific testimony in trials is largely guided by the Daubert standard. The UK Law Commission interprets the four Daubert principles as (The Law Commission, 2009):

- ordinarily a key question is whether the theory or technique in question can be (and has been) tested;

- a further pertinent consideration is whether the theory or technique has been subjected to peer review and publication;

- in the case of a particular scientific technique, the court should ordinarily consider the known or potential rate of error and the existence and maintenance of standards controlling the technique's operation; and

- widespread acceptance can be an important factor in ruling particular evidence admissible, and a known technique which has been able to attract only minimal support within the relevant scientific community may properly be viewed with scepticism.

The standard places responsibility for assessing the reliability of testimony with the trial judge. The extent to which the method employed to produce evidence conforms to the principles of scientific method is used to guide judgements of acceptability. The conduct of forensic investigations in cloud environments in both the United States and United Kingdom will presumably be subject to the same tests, if the resulting evidence is to be admissible in court.

The empirical testing of cloud forensic methods may be challenging due to the rapidly evolving nature of the technology. Empirical testing of forensic tools typically employs standard data sets (Guo, Slay, & Beckett, 2009), but it is unclear how these could be developed for cloud forensic methods. Certainly, there is a clear need to develop a standard evaluation method and data set for cloud forensics, if results of cloud forensic investigations are to pass the Daubert principles. These criteria affect not only cloud investigations but traditional computer forensics

investigation as well. As Marsico notes:

> ...the court does not have a true universally accepted method to rely upon. To further complicate the problem, many self-proclaimed computer forensics experts take what they feel are the best aspects of several approaches and create their own methodology. (Marsico, 2004)

Expert witnesses could be faced with the additional challenge of having to explain the concept of cloud computing to a jury. It must be remembered that juries in common law systems are made up of individuals from the general-public, very often, people who only use a personal computer to perform simple tasks. It can be expected that before a judge can allow a jury to listen to evidence retrieved from the cloud, they must understand what a 'cloud' is, and how it works. This could further prolong court proceedings and expert witnesses will be faced with the daunting task of ensuring juries fully understand the concept of the cloud.

The evolution of cloud computing forensics is in its infancy. Currently there is not a standard method or tool set for conducting cloud investigations, or even for evaluating and certifying proposed tools. The presentation of evidence derived from a cloud service will likely be problematic in the near future.

## RELATED WORK

The potential benefits and challenges of cloud computing for digital forensic investigations have been discussed by several authors (Biggs & Vidalis, 2009, 2010; Reilly, Wren, & Berry, 2010; Ruan, Baggili, Carthy, & Kechadi, 2011; Ruan, Carthy, Kechadi, & Crosbie, 2011; Taylor, Haggerty, Gresty, & Hegarty, 2010; Wolthusen, 2009). Reilly et al. (2010) speculate that one potential benefit of cloud computing is having data in a centralized location, which can mean incidents can be investigated more quickly. Wolthusen (2009) notes that when attempting to locate evidence in a distributed environment such as the cloud, major challenges will need to be overcome because evidence could be located across several locations making evidence collection difficult. The distribution of evidence can be across multiple virtual hosts, physical machines, data centres and geographical and legal jurisdictions. The distributed nature of control and storage in a cloud (and the ephemeral nature of virtual instances) will also likely make tracing activity and re-construction of events more challenging (Wolthusen, 2009).

Other challenges identified includes a loss of important forensic information such as registry entries (on Microsoft Windows platforms) temporary files, and metadata which could be stored in the cloud as well as a lack of tools for dealing with investigations involving cloud data centres (Taylor, et al., 2010). As part of the Cloud Computing and The Impact on Digital Forensic Investigations (CLOIDIFIN) project, Biggs and Vidalis (2009) reported that very few High Tech Crime Units (HTCUs) in the UK were prepared to deal with crimes involving cloud computing. Even when HTCUs are prepared to investigate such crimes, current legislation for accepting digital evidence in court presents further challenges (Biggs & Vidalis, 2009).

Ruan et al. (2011) define cloud computing forensics as a form of network forensics, arguing that cloud environments are essentially a form of public and/or private computer network. However, this definition does not incorporate the virtualized nature of clouds, which is likely to have a significant impact on forensic investigations beyond the networked aspect. The paper proceeds to sketch a process model for forensic investigation in a cloud environment; however, it is unclear how the model is to be evaluated. Ruan et al. (2011) reported on a survey of digital forensic practitioners. The survey was conducted with the intention of establishing the views of practitioners concerning the impact of cloud computing on future digital forensic investigations. The results of the survey indicate considerable diversity of opinion within the practitioners surveyed. For example, little agreement was reached as to the definition of cloud forensics.

Taylor et al. (2010) have extensively examined the legal issues surrounding evidence retrieved from a cloud environment. Just as in a traditional investigation, any evidence gathered from the cloud should be conducted within local laws and legislation. For example, in the United Kingdom, the Data Protection Act 1998, the Computer Misuse Act 1990, as well as the Criminal Procedure and Investigations Act 1996 and the Criminal Justice Act 2003 will also apply to cloud computing investigations (Taylor, et al., 2010).

Reilly et al. (2010) and Roussev et al. (2009) have suggested cloud computing environments as a basis for *conducting* forensic investigations. Dedicated virtual instances can be ready and waiting on 'stand-by' to assist in gathering evidence from an incident or crime. Cloud storage can be used to store images gathered from investigations, and the extensive computing resources available can also be used to perform brute-force cracking attacks on passwords and encryption keys (Reilly, et al., 2010). Roussev et al. (2009) have proposed the use of cloud computing as a means of 'speeding up' forensics investigations. A framework which takes advantage of distributed computing resources was developed and results showed that such an environment could assist investigators examine large forensic data sets in real time (Roussev, et al., 2009).

## FUTURE WORK

As discussed in previous sections, current issues in the area of cloud forensics investigations support the development of an immediate research agenda in the area of procedures, tools, methodologies, and specific environments. These issues will be of concern to both the public and private sectors. This section specifically examines several areas of research that the authors intend to conduct to further understand digital forensics investigations in the cloud that include: an analysis of cloud service usage, the effectiveness of acquisition methods, an understanding of commercial cloud environments, an investigation of cloud forensic management, and the impact of the cloud on mobile devices.

### Analysis of Cloud Service Usage

Alternatives to traditional storage are becoming commonplace in increasingly networked societies. Cloud service providers such as Amazon, Google and Dropbox are offering alternatives to traditional file storage, email and collaboration solutions. As these options continue to become available, the likelihood of these environments being investigated increases. Initially, this research needs to determine the number of organizations that are utilizing these environments. This could be achieved through a targeted in-depth structured survey with organizations and through the implementation of broader Web-based surveys. When an understanding of how these environments are being utilized has been achieved, then the research should examine the existing policies and procedures associated with these environments. Do existing policies and procedures address any issues associated with digital forensics and the investigation of these environments?

The second stage of this survey should investigate private sector need for support from police. In doing so, it will need to establish the amount of collaboration and support currently provided by local, regional and national police. This survey needs to not only investigate existing support but examine ideas on improvements or the removal of existing support where it is deemed inefficient or unnecessary. Again, this can be accomplished through a combination of in-depth structured surveys and broader Web based surveys.

## Acquisition Methods for Cloud Environments

The third stage in the investigation needs to evaluate current methods of evidence acquisition from the cloud. This includes examining effective ways to capture data from the client and the service provider. Will accepted forensic imaging tools effectively capture data from the cloud? If current tools are deemed insufficient, can they be modified to capture data or will new tools need to be developed to achieve this task? If new tools need to be developed what are the application development requirements?

A cloud test-bed can be used to mimic the main cloud services (SaaS, PaaS or IaaS) offered by cloud providers. The authors believe each of these services will require a unique acquisition methodology and each service will provide a unique set of challenges that need to be overcome. For example, in an IaaS environment an acquisition methodology needs to be developed such that forensic 'images' can be acquired from the virtual machines running in this environment. Each cloud environment could be configured to address known environmental and trust issues that exist in these environments. For example, a private cloud located in a financial organization is going to have more control over the environment than users of a public cloud. Hence, the trust issues associated with user rights, operating system functionality and capability restrictions can be minimized in a corporate environment. On the other hand, a public cloud that allows end-users to utilize multiple operating systems, possesses administrative rights, provides minimal audit information, and introduces more complexity and risk into the environment, which makes the forensic acquisition of these environments more challenging. These environments raise several questions. Can the cloud be suitably stabilized to allow investigators to take an accurate representation of the evidence at a specific point in time? Can datasets be developed to effectively mimic this environment? With the investigator not having complete control of the environment, can the investigator be sure evidence is not in the process of being altered in the cloud at that moment in time?

To complicate the issue, cloud storage providers employing a distributed file-system might mandate the development of alternative tools to recognize that such an environment is being used and effectively 'fetch' all the evidence from various physical locations. There is also the problem of encryption that needs to be overcome such that acquired evidence is not effectively a forensic 'image' of encrypted data.

Once the data capture issues have been resolved, then the type of useful data that can be captured on both the client and the service providers needs to be identified. This would enable investigators to concentrate requests for data from cloud providers on known artefacts. Evidence gathered in this way could conceivably contribute to a revised or completely new cloud investigation process model.

## Commercial Cloud Providers

The fourth stage of the investigation expands the scope of the research to identify potential residual artefacts that are currently left by commercial cloud providers. Individual providers will be contacted to inquire about potential collaborations in the forensics analysis of data residing on public clouds. An example of this stage of the research would be an investigation into Google Apps. The client would need to be extensively examined to thoroughly understand and identify the footprint it leaves on the operating system. The investigation can then focus on how the client stores information. In other words, does it cache a copy of emails or any type of document on the hard disk drive? During the experiment, the client can be used to connect to Google Apps for a period of time. Documents received through Google Apps can be viewed online without downloading local copies to the hard disk. The hard disk of the client computer can then be imaged using traditional resources and the data examined. Some of the questions that could be investigated include:

- whether it is possible for an investigator to recover email messages including the email body and header information;

- the availability of log files for recovery;

- the recovery of documents viewed using Google Docs; and

- how principles of forensic investigation can be applied (i.e. so that evidence is not altered during the course of an investigation).

The experiment then could be expanded to other cloud-based storage solutions such as Dropbox[xi], Wuala[xii], Syncplicity[xiii] and SpiderOak[xiv]. A data set consisting of a selection of files would need to be stored in the cloud using the above solutions. Information gathered from the first stage of the experiments can be utilized to establish typical activity in cloud environments.
A detailed experimental design would need to be developed consisting of a number of file manipulations, such as storing, moving and deleting files in order to mimic real world activity by a suspect in the cloud. The investigation will help to determine data recoverability from client-side computers along with identification of potential artefacts created from file manipulation. This experiment can also be extended to investigate the clouds' effect on digital evidence timestamps. The reality is that the cloud client and cloud storage server could reside in different time zones. By observing the time when manipulations are made to the data set during the course of the experiment, the timestamps can be examined and any affects the cloud has on these timestamps can be noted.

## Cloud Forensic Management

The idea of using the cloud to host a 'forensic server' to be used to conduct investigations has already been proposed by Reilly et al. (2010). Several issues are raised by this prospect, including, for example, the forensically sound transfer of evidence from the source of the investigation to the cloud storage. Such a prospect also raises challenges concerning the management of the chain of custody for evidence, and the consequences of a security breach suffered by the cloud provider.

Another area of research that is directly relevant to cloud forensics is the growing problem of how to handle large data sets (Tabona & Glisson, 2011). With evidence from the cloud expected to be much larger than what current investigators are examining presently, how do investigators process the evidence effectively along with storing this evidence safely and securely? Does the solution for large-scale cloud data processing hashing techniques, customized information retrieval solutions, random data sampling, parallel processing solutions or some combination of these solutions? One solution for storage could be the use of the cloud itself.

Issues, which can be examined, include:

- ensuring the evidence is unaltered when transferred to and from as well as when stored in the cloud;

- ensure that by using the cloud, local laws, such as data protection, are observed if the cloud is used to store evidence;

- enforcing environmental and evidential integrity to eliminate unauthorized access along with intentional and unintentional modification or deletion of evidence;

- in the event where sensitive evidence (explicit images and videos) are stored in the cloud, access to this evidence is limited and other users of the cloud do not have access to this material.
- is it possible to use the computing power of the cloud to augment the process of data intensive cases or to improve password cracking capabilities?

Mechanisms to prevent the above, such as encrypting the evidence, could be implemented as a potential solution. However, due to the large size of the evidence, this could be a cumbersome resolution. Therefore, research should be conducted to determine the most effective and efficient security solutions for the cloud. The idea is to provide investigators with appropriate data so that they can implement the appropriate level of security for their environment.

## Mobile Cloud Device Environments

Coupling increased mobile phone subscriptions with growth in the smart phone markets indicates that these devices are continuing to become an integral part of technologically advanced societies. With this assimilation, it is only a matter of time before these devices are used in conjunction with cloud services. The same surveys to establish use in the corporate and private sectors of society need to be implemented in reference to mobile phones and other mobile devices like tablets. How many organizations and individuals are currently utilizing cloud services on mobile devices? From an organizational perspective, what is the perceived risk and have these risk been effectively mitigated through appropriate policies and procedures? These devices largely have proprietary operating systems. Experiments need to be designed and executed to determine if any residual artefacts remain on the devices when they interact with cloud services. Test data sets will need to be developed and implemented to not only test the residual artefacts, but also set performance benchmarks for assessing commercial solutions. The residual information could plausibly change with the release of every operating system, from the various providers, magnifying and complicating this topic. The impact of the use of encryption keys to secure cloud access and the storage of these key either on the device or with a third party needs to be examined as well. The effectiveness of sandboxing memory areas in mobile devices used in conjunction with cloud services is another area that needs to be investigated.

## CONCLUSION

This paper has argued that conventional methods and guidelines suggested for conducting digital forensics could well be insufficient in a cloud environment. If current forecasts are correct, more businesses and organizations will be moving their data to cloud environments. Together with a continued growth in cyber-crime, this transition could mean there will soon be a demand to conduct forensics investigations in such environments. Such investigations would currently be hampered due to the lack of guidance concerning methods and software tools to retrieve evidence in a forensically sound manner. There is also the need for legal issues regarding clouds including data retention and privacy laws to be re-examined, following the widespread adoption of cloud technologies. Finally, there is also the need for the digital forensics community to begin establishing standard empirical mechanisms to evaluate frameworks, procedures and software tools for use in a cloud environment. Only when research has been conducted to show the true impact of the cloud on digital forensics, can we be sure how to alter and develop alternative frameworks and guidelines as well as tools to combat cyber-crime in the cloud.

---

[i] Google Apps - Web-based email, calendar, and documents for teams. Retrieved February 2, 2012, from http://www.google.com/apps/intl/en/business/index.html

[ii] Google App Engine. Retrieved February 2, 2012, from http://appengine.google.com

[iii] Amazon SimpleDB. Retrieved February 2, 2012, from http://aws.amazon.com/simpledb

[iv] Amazon Elastic Compute Cloud (Amazon EC2). Retrieved February 2, 2012, from http://aws.amazon.com/ec2

[v] AccessData FTK Imager. Retrieved February 2, 2012, from http://accessdata.com/support/adownloads#FTKImager

[vi] Coreutils - GNU core utilities. Retrieved February 2, 2012, from http://www.gnu.org/software/coreutils/

[vii] SpiderOak. Retrieved February 2, 2012, from http://www.spideroak.com

[viii] AccessData Forensic Toolkit (FTK). Retrieved February 2, 2012, from http://accessdata.com/products/computer-forensics/ftk